# Classes in Object-Oriented Modeling (UML): Further Understanding and Abstraction

Sabah Al-Fedaghi

*sabah.alfedaghi@ku.edu.kw, salfedaghi@yahoo.com*
Computer Engineering Department, Kuwait University, Kuwait

**Summary**

Object orientation has become the predominant paradigm for conceptual modeling (e.g., UML), where the notions of class and object form the primitive building blocks of thought. Classes act as templates for objects that have attributes and methods (actions). The modeled systems are not even necessarily software systems: They can be human and artificial systems of many different kinds (e.g., teaching and learning systems). The UML class diagram is described as a central component of model-driven software development. It is the most common diagram in object-oriented models and used to model the static design view of a system. Objects both carry data and execute actions. According to some authorities in modeling, a certain degree of difficulty exists in understanding the semantics of these notions in UML class diagrams. Some researchers claim class diagrams have limited use for conceptual analysis and that they are best used for logical design. Performing conceptual analysis should not concern the ways facts are grouped into structures. Whether a fact will end up in the design as an attribute is not a conceptual issue. UML leads to drilling down into physical design details (e.g., private/public attributes, encapsulated operations, and navigating direction of an association). This paper is a venture to further the understanding of object-orientated concepts as exemplified in UML with the aim of developing a broad comprehension of conceptual modeling fundamentals. Thinging machine (TM) modeling is a new modeling language employed in such an undertaking. TM modeling interlaces structure (components) and actionality where actions infiltrate the attributes as much as the classes. Although space limitations affect some aspects of the class diagram, the concluding assessment of this study reveals the class description is a kind of shorthand for a richer sematic TM construct.

*Key words:*
*Conceptual analysis, logical design, classes, static model, behavioral model*

## 1. Introduction

Modeling in software engineering and system engineering involves the process of collecting and analyzing information about a system to build a representation of the involved domain. A system is "what is distinguished as a system" [1] carved out of reality. Distinguishing an entity as being a system is a necessary and sufficient criterion for it being a system [1]. The internality of a system includes structural and behavioral aspects that form a single coherent, distinguishable whole. The underlying assumption of this line of thinking is that reality embeds domains that are susceptible to being expressed as formalized or semi-formalized models. Conceptual modeling is concerned with identifying, analyzing, and describing the essential concepts and constraints of such domains [2]. Conceptual models are built in the early stages of system development, preceding design and implementation; they can also be useful, even if no system is contemplated, to clarify ideas about structure and functions in a perceived part of the world [3]. Wand et al. [4] state four purposes for conceptual models: supporting an analyst's understanding of an application domain, communicating with stakeholders, communicating with implementers, and documenting system rationale for future needs.

Currently, the object-oriented paradigm (e.g., UML) has become prevalent for conceptual modeling [5]. Object-orientation allows computer scientists to make compelling comparisons to the real world, where "all tangible things are objects, and where it is not hard to conceive of most intangible things as objects, too. With object-oriented programming, computer science takes a break from mathematics, and is influenced by philosophy" [6].

1.1 Focus: The Class Diagram

Object-orientation is based on objects and classes as the primitive building blocks. According to Pedroni and Meyer [7], the concepts of classes, message passing, and single and multiple inheritance were initially programming concepts, but they are in fact useful for a far more general purpose: designing systems, modeling systems, and more generally thinking about systems. The modeled systems are not even necessarily software systems: They can be human and artificial systems of many different kinds (e.g., teaching and learning activities) [7].

The UML class diagram is called a "bridge" between software specification and software realization. It is the most common method in modeling object-oriented systems [8] and used to model the static design view of a system. It is described as a central component for representing a domain in a platform-independent manner and serves as a basis for generation of platform-specific details that are required for further generation of a software code [9]. During the requirements analysis, the class diagram is viewed from the conceptual perspective and is used as a problem domain dictionary (potential classes), and it contains least specific notation [10].

### 1.2 Research Problem: Difficulties in Class Semantics

However, in spite of the wide adaptation of the object-oriented approach and UML as the most common modeling paradigm, "The use of object-concepts in conceptual modeling has not been widely adapted. A main reason is that there are no generally accepted semantics of these concepts as conceptual modeling elements" [4]. In object-oriented modeling, "The basic concepts are tightly interrelated and cannot be easily taught and learned in isolation" [11]. This complexity is intrinsic to object orientation and cannot be removed [7]. For example, in modeling the relationship between notions of species, for apes and particular apes, we say that there is a concept SPECIES (representing the set of all species), with instances such as APE. However, APE may be viewed as a set of all apes. It may be argued that APE may be modeled as a subconcept of SPECIES. However, since APE is a set of all apes, SPECIES, being a super concept of APE, must contain all apes as their members, which is clearly wrong [5][12].

In general, according to Sedrakyan et al. [13], "There is a certain degree of difficulty in understanding a system represented by means of UML diagrams." A survey of UML practitioners [14] [9] shows class diagrams are not fully used for further software development, either for code generation or documentation. Hence, class diagram has lost the role it could have played in software development (i.e., serving as a bridge between system specification on the user side and software components on the developer side) [15][9]. It is reported that some commercial industries find modeling cumbersome and slows down productivity [16] [9]. "For such projects, it makes sense to use UML as a sketch and have your model contain some architectural diagrams and a few class and sequence diagrams to illustrate key points" [17] [9].

In Halpin's [18] opinion, class diagrams "have limited use for conceptual analysis and are best used for logical design." According Halpin [18],

> When I'm performing conceptual analysis... I sure don't want to bother about how facts are grouped into structures. Whether some fact will end up in the design as an attribute is not a conceptual issue. UML covers both data and behavioral modeling, and lets you drill down into physical design details. You can declare whether an attribute is private, public, or protected, what operations are encapsulated in an object, and whether an association can be navigated in one direction only.

Hay [19] argued, "There is no such thing as 'object-oriented analysis' only object-oriented design" and that "UML is … not suitable for analyzing business requirements in cooperation with business people." The UML model is complicated and much harder to present to an audience of business people. UML and object-oriented analysis are fundamentally design tools, but not ones suitable for analyzing business requirements in cooperation with business people [19].

### 1.3 Objectives

This paper is an attempt to understand object-oriented modeling further as exemplified by the class concept in UML. Class diagrams convey rather little semantics on their own [20]. To address the issue of class concept semantics, we use a thinging machine (TM) model that provides a parallel conceptual representation for classes. The main TM construct is called a *thimac* (i.e., a *thi*ng/*mac*hine). We start with the simplest class structure, hoping to expand the analysis if it proves fruitful. The research strategy is to "translate" classes into TM notions then observe the differences, especially with exposed features in both types of representations.

### 1.4 Overview

The next section provides a review and example of TM modeling. The review aim is to achieve a self-contained paper while the example is a new contribution. Section 3 provides two samples of re-modeling of single UML classes. Section 4 presents a bank account class diagram with two subclasses. Section 5 gives further analysis using instances of classes.

## 2. TM Modeling

TM modeling articulates ontology in terms of an entity that is simultaneously a *thing* and a *machine*, called a *thimac* [21-31]. A thimac is like a double-sided coin. One side of the coin exhibits the characterizations assumed by the thimac, whereas, on the other side, operational processes emerge that provide dynamics (facilitation of change in thimacs). A thing is simple unity of "what is there," of what stands by itself subjected to doing. By contrast, a machine is "what it does" as mere actionality (there is only *in potentia*; i.e., as essential possibilities rather than as substantive particular realities); that is, a machine is "turned on" by time flow that gives rise to actions. A thimac could be a composite entity of nets of subthimacs. To complement Gaines's [1] definition of a system mentioned in the introduction, a system is a thimac that is the totality of a hierarchy of thimacs.

In the object-oriented models (e.g., UML), this two-faceted construct is expressed as a hierarchy of classes structured by the classification link. Concepts are viewed alternatively as objects and as classes of their subconcepts [32]. According to Pirotte and Massart [32], "Each concept is a two-faceted construct with an object facet (a concept is an instance of a more abstract concept at the next higher level of the taxonomy) and a class facet (a concept is a class of refined concepts that are its instances at the next lower level)."

The TM is the plurality factor by which a thing becomes unified. For example, a water machine *processes* (fusion process) O and $H_2$ to *create* water. The thimac is a template (class-like) for such fusion that is realized (object-ed) when the thimac is eventized with a time subthimac. In TM modeling, thimacs are the "alphabet of being" [33]. Interaction among thimacs involves flows of things and triggering, as will be illustrated later.

## 2.1 Genetic Actions

Thimacs are the source for generic actions to be applied in conceptual modeling to describe structure/behavior as a world of systems (thimacs). The generic (i.e., not made up of more basic operations) actions in the machine (see Fig. 1) are merely potentialities that do not yet act, waiting to be eventized, which will be discussed when modeling system behavior. The actions represent the kind of compresent action in which all things essentially engage.

Fig. 1 can be described as follows:

**Arrive:** A thing moves to a machine.

**Accept:** A thing enters the machine. For simplification, we assume that all arriving things are accepted; hence, we can combine the arrive and accept stages into one stage: the **receive** stage.

**Release:** A thing is ready for transfer outside the machine.

**Process:** A thing is changed, but no new thing results.

**Create:** A new thing is born in the machine.

**Transfer:** A thing is input into or output from a machine.

Additionally, the TM model includes memory organization that plays the role of storage for each action. For simplification purposes, one may assume each thimac has a single storage area. Additionally, the TM model includes the mechanism of triggering (denoted by a dashed arrow in this study's figures), which initiates a flow from one machine to another. Multiple machines can interact with each other through movement of things or triggering. Triggering is a transformation from one series of movements to another.

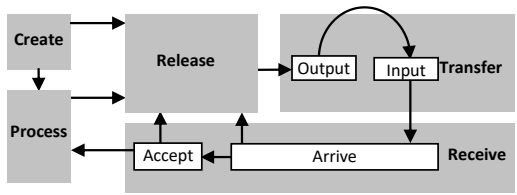

Fig. 1 The thinging machine.

## 2.2 Example

Haugen et al. [34] model cooking beef in the sequence diagram shown partially in Fig. 2. According to Haugen et al. [34], from the figure, the work of the cook making beef is assumed to be intuitive, as follows: The cook receives an order for the main dish and then turns on the heat and waits until the heat is adequate. Then he fetches the sirloin meat from the refrigerator before putting it on the grill. Then he fetches the sirloin from the stove. He then sends the steak to the customer.

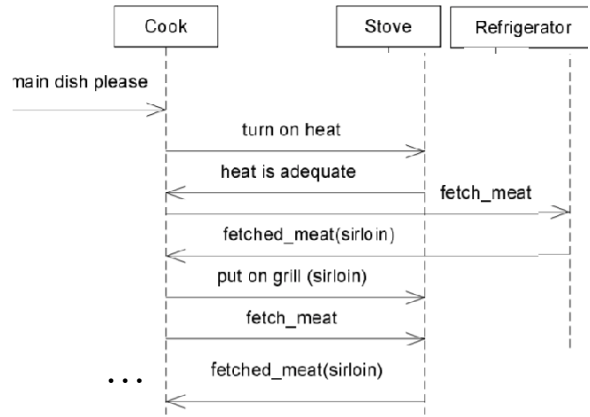

Fig. 2 Partial view of the sequence diagram that models making beef (from Haugen et al. [34]).

### 2.2.1 Static TM Model

Fig. 3 shows the TM model that corresponds to this sequence diagram. In the figure, the customer (circle 1) creates an order (2) that flows (3) to the cook (4), who processes it (5) to trigger (6) the heat in the stove (7). The ON state is processed (8) to the right temperature. The sirloin (9) is fetched from the refrigerator (10) by the cook (11) to be put on the grill (12). There, it is processed (13) to trigger the creation of a steak (14), which flows to the customer (15 and 16).

The TM description in Fig. 3 is static in the atemporal sense. It involves the *spatiality* of things' boundaries, such as customer, cook, stove, refrigerator, and grill pan, as given in Haugen et al.'s [34] description. It also involves *actionality*, the five TM generic actions, and flow/triggering (arrows). The actions are not PROCESSES (in the generally understood meaning). For example, *create* is *what the thimac machine does*. *Create* is a noun that refers to the potential act of creation, not a verb that indicates a PROCESS in time. One may say the *create* stage should have been named the *creation* stage. However, as we will see when specifying the dynamic model from the static model, this *creation*, when time is involved, should become the event *create*. Accordingly, we have opted to use *create* on both static and dynamic levels. Similar discussion can be applied to the other TM actions.

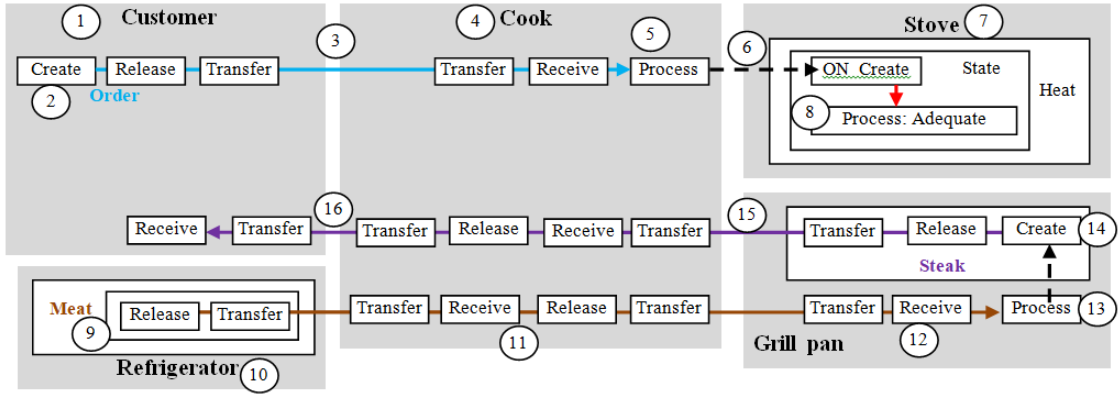

Fig. 3 The static model of making beef.

### 2.2.2 TM Events Model

In TM modeling, actionality is a static notion embedding potentiality of events and behavior that appears when time is added to the static model. A thimac in the static model "exists/appears" in the system as a thing and as a machine, but without "behavior" (e.g., time-oriented notion). The static model gains behavior through events. An event is formed from
- A thing (has specific spatiality; e.g., a boundary) and a machine (has actionality)
- Time.

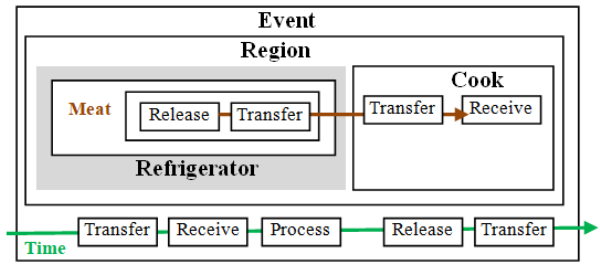

Fig. 4 The event *The cook fetches meat from the refrigerator.*

For example, Fig. 4 show the event *The cook fetches the meat from the refrigerator*. The region in the figure is a subdiagram of the static model. To develop the events model, it is necessary to identify the underlying *decompositions* (regions) where behavior can happen (potentiality of dynamism), as shown in Fig. 5. Accordingly, in Fig. 5, the following event regions are developed.

Event 1 ($E_1$): The customer creates an order that flows to the cook.
Event 2 ($E_2$): The cook processes the order and turns ON the stove.
Event 3 ($E_3$): The heat in the stove reaches an adequate level.
Event 4 ($E_4$): The cook fetches the meat from the refrigerator.
Event 5 ($E_5$): The cook puts the meat on the grill.
Event 6 ($E_6$): The meat is processed, thus creating a steak.
Event 7 ($E_7$): The cook takes the steak from the grill.
Event 8 ($E_8$): The cook sends the steak to the customer.

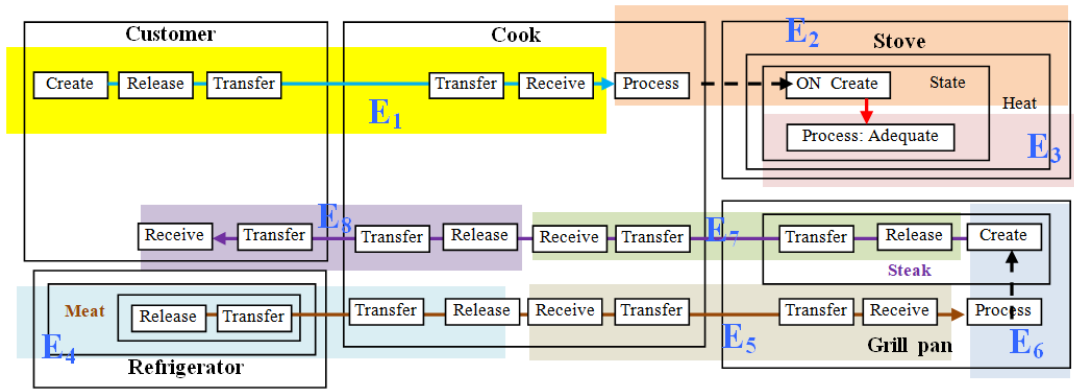

Fig. 5 The events model.

### 2.2.3 Behavioral model

The specification of events in Fig. 5 requires defining the chronology of legitimate events. To accomplish that, we follow Ehrich and Goguen's [35] metaphor of the "blinking observer." According to Ehrich and Goguen [35], assume that you are an observer who is always blinking. Then, when you look at a thing, you will see its traces of events and values as follows: Each time you open your eyes, you will take note of the events happening at that time. You will see all the traces of all the events and will notice which events happen at the same time (synchronization). Events may appear interleaved and/or simultaneous. Fig. 6 shows this registering of the chronology of events, thus defining the behavioral model where the legal sequence of events for an order is specified.

## 3. Remodeling a Single Class

In object-oriented modeling, an attribute is any member of a class of entities that is capable of being attributed to objects. Terms that are similar attributes include property characteristic, type, and predicate. One of the disagreements in ontological research concerns the rule that states intrinsic attributes and associations should never be modeled as entity types in a conceptual model [36]. In modern philosophy, several debates involve the fundamental nature of attributes. Plato called them "forms" and viewed them as *universals*; that is, as capable of being instantiated by different objects (entities that can have instances).

TM modeling provides a representation without a sharp distinction between classes and attributes, as in the case of UML. A class, a subclass, and attributes are all thimacs or subthimacs. We show that such an approach takes the notion of encapsulation of structure and activities to its end, where "methods" are wrapped with "attributes."

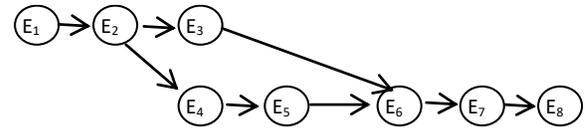

Fig. 6  The behavioral model in terms of the chronology of events.

In programming, "encapsulation" means data and programming methods are wrapped together in an object. At the conceptual level, such a notion is translated as wrapping attributes and methods of the class/object. In TM modeling, because classes and attributes are all thimacs (machines), they both have wrapped structures and actions.

### 3.1 Example 1

Fig. 7 illustrates these notions for a single class, *Person*, taken from [37]. Fig. 8 shows the TM representation of this *Person* class. Structurally, the so-called attributes are subthimacs located within the thimac containing *Person*. Each of these "attributes" has its own events; for example, getName () is an event that gets the particular value of Name (release+transfer) at the global level and feeds it (transfer+receive) to Name. Fig. 9 shows the difference between the static (passive) thimac setName() and the (active) event setName().

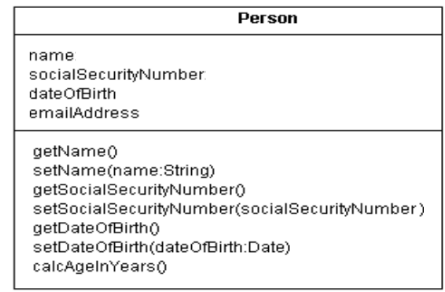

Fig. 7  The class *Person* (incomplete, adopted from [37]).

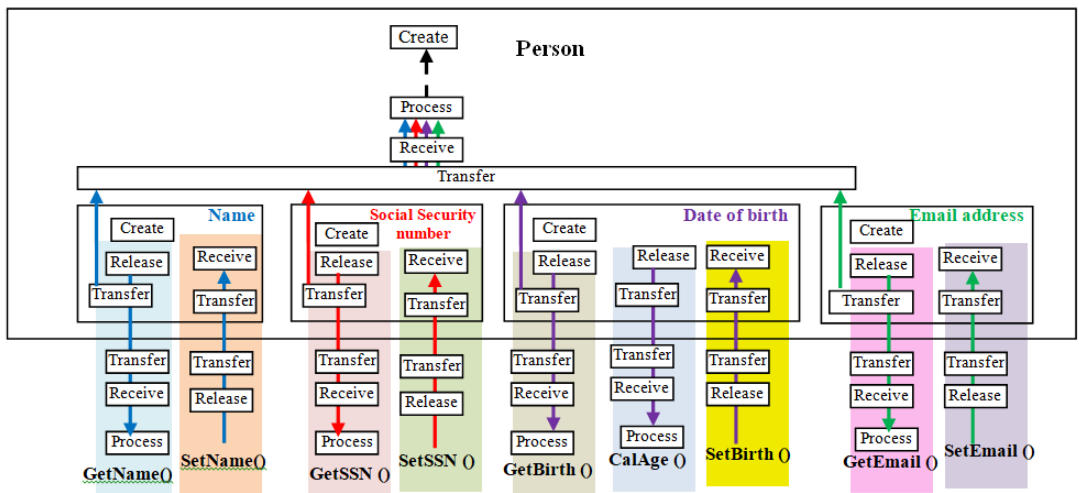

Fig. 8  The TM representation of the class *Person*.

As an illustration of these passive/active features, consider a stack of integers as a machine where we observe the following specifications and events of "push (9)" in a semiformal language.

Passive:

*stack.transfer()→receive()---> stack.state.create(**empty**)*

Active:

*event[stack.transfer(**9**)→receive(**9**)] ---> event[stack.state.create(**not empty**)]*

Note that "push 9" is implemented as transfer and receive TM actions. In addition, note that actions (typically called atomic operations in the terminology of the object-oriented community [35]) become events only if time is involved.

Note the TM feature of a complete wrapping of structure and behavior in every thimac and subthimac without the rigidity of the UML class template. As an illustration, we show the constructing of *Person* from its "attributes," say, as a record or tuple in database terminology. No sharp distinction of attributes and methods exists, as in the UML class. The methods are completely dissipated into the TM actions. There is no need for the mysterious notion of UML operations where objects *operate on* or *are operated on*.

The notion of *state* (of an object) in TM modeling exists at a different level of specification that involves events/time. Thus, the notion of behavior appears at this second level of specification. The methods are constructed in terms of the five actions. See Fig. 9 for a sample "attribute" name that shows the two levels of specification.

The class template seems to be a shorthand notation for the TM representation. This supports Knapp's [20] thesis that class diagrams convey little semantics on their own. Additionally, the TM representation as a high-level conceptual description supports Halpin's view [18] that class diagrams have limited use for conceptual analysis and are best used for logical design. The class diagram furnishes a specific design setting of a structure—in Halpin's words, "how facts are grouped into structures" [18].

The class diagram can be generated from the TM model. If we eliminate TM actions and arrows, then we produce a very similar notation to that of the usual definition of the class (see Fig. 10). Of course, in this case, we have to introduce the notion of operation that mixes actionality and events or behavior.

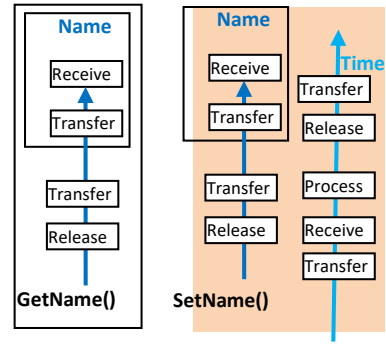

Fig. 9 The thimac setName() (left) and the event setName (right).

### 3.2 Example 2

According to Steinhart [38], the object-oriented approach provides a hierarchy of classes, much like the Aristotelian genus–species hierarchy. TM modeling provides a hierarchy of thimacs—one that defines classes and attributes. Thimacs have subthimacs. For example, one might define the class *Human* by giving it instance variables of name, weight, and gender [38]. In TM modeling, the thimac *Human* has the subthimacs name, weight, and gender (see Fig. 11). Fig. 11 is a static description that includes the five generic actions. The actions here are potentiality for behavior. These actions are "activated" in a time or event context.

Then one might define the instance of *Human* by giving it the name "Bob," a weight of 150 units, and the gender "male" (see Fig. 12). For example, eating is one of the processes of the instance labeled *Human*. Hence, *eat* is a method attributed to a *Human* thimac [38] (see Fig. 13).

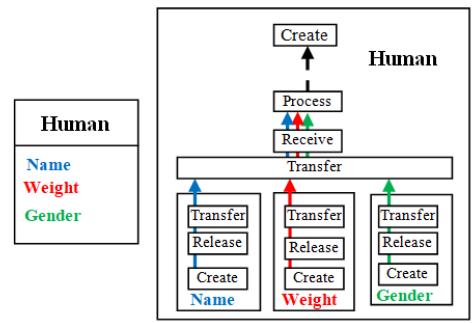

Fig. 11 Object-oriented *human* class (left) and the corresponding TM thimac (right).

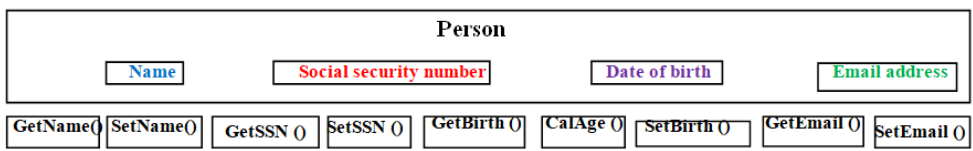

Fig. 10 Class structure produced by eliminating actions and arrows in the TM representation.

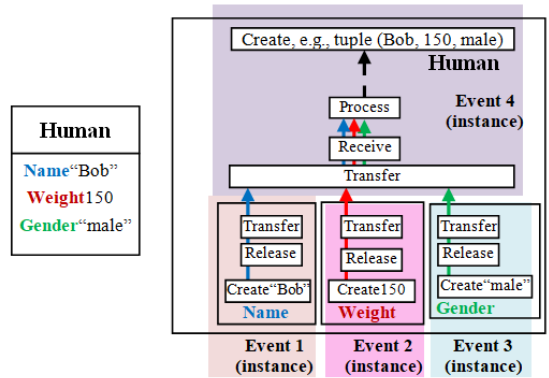

Fig. 12 Object-oriented human instances (left) and the corresponding TM instances (right).

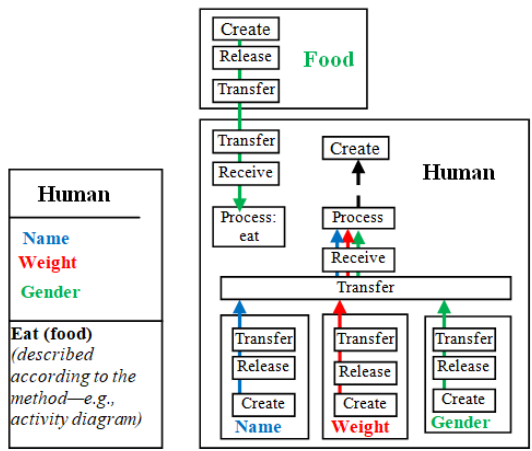

Fig. 13 Object-oriented human class with a method (left) and the corresponding TM thimac (right).

Then one might define two instances of *Human*, giving the first the name "Bob," the weight 150, and the gender "male" and the second the name "Sue," the weight 110, and the gender "female." Processes associated with the instances of a class are defined by the methods of the class. For example, eating is one of the processes of an instance labeled *Human*. Hence, *eat* is a method used by a *Human* thimac; when an instance of *Human* eats, its weight is increased. In continuing this example, one might define a class Food.

## 4. A Class with Two Subclasses

According to Donald Bell in *IBM Developer* [39] (referring to the Oracle and Java tutorials [40]), there are two basic categories of diagrams in UML 2: structure diagrams and behavior diagrams. The structure diagrams include the foundational class diagram that provides an initial set of notation elements that all other structure diagrams use. *Inheritance* in object-oriented design refers to the ability of one class to *inherit* the identical functionality of another class and then add new functionality of its own. Fig. 14 shows an example of how both CheckingAccount and SavingsAccount classes inherit attributes from the BankAccount class [39].

Fig. 15 shows the TM model that corresponds to this bank account class diagram. The general structure is formed from the bank account (1 [pink numbers]), owner (2), checking account (3), and savings account (4), with balance (5) divided into a checking balance (6) and a saving balance (7) and respective amounts (8 and 9) that are repeated for the purpose of diagrammatic convenience. Note that the checking account (3) and the savings account (4) (rectangles with thick circumferences) are mirror images of each other; hence, we will detail only the savings account (4).

In a typical transaction, first, bank accounts (1) are activated (processed) (10). Note that representation of the bank accounts (many of them denoted by "…" in the top right corner of the figure) is simplified by eliminating the step of progressing from the set of bank accounts to a single bank account. This simplification is performed to go along with the given UML diagram that starts with a single bank account.

Processing the bank account (e.g., by displaying a starting page) is followed by processing the owner of the account through requesting and receiving his or her identification (11 and 12).

We assume this is followed by selecting (processing [14]) the savings account.

Inside the savings account are the withdrawal (15) and deposit (16) thimacs (note the process states in both of these thimacs).

- Accordingly, the amount of the transaction is inputted (17).
- The amount flows (19) are to be compared with the current balance (20 and 21) that calculates the result of the deposit or withdrawal (22).
- If "deposit" was selected for the transaction, then the new balance will flow to the savings account balance (23).
- If "withdrawal" was selected, then the result will flow to be processed (24).
  - If the result is negative, then an insufficient fund message is sent (25).
  - If the result is positive, then a "give" instruction is issued (26) that triggers (27) the release of money, resulting in a new balance.

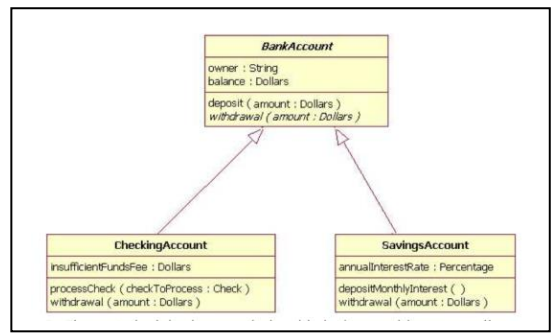

Fig. 14 The class diagram given in [39].

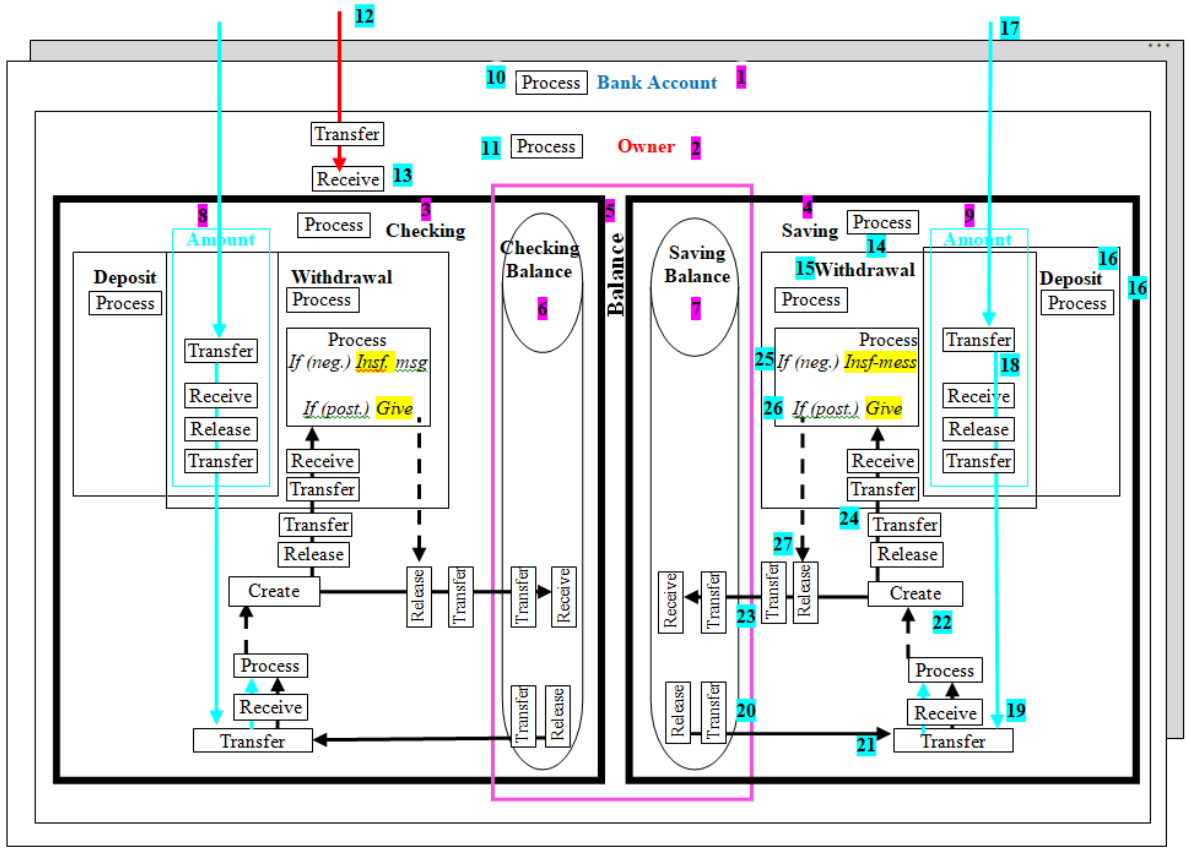

Fig. 15 The TM model of the class diagram of the bank account.

Note the interlacement of structure (components) and actionality (five generic TM actions) in the model of Fig. 15. Still, the model is static and the dynamic aspects will next be superimposed on this initial model. Actions infiltrate the attributes as much as the classes. The TM model presents a foundation for modeling the organizational framework of the system that involves classes. We observe that the UML class description is a kind of shorthand for this TM foundation. Thus, to reproduce the UML class diagram from the TM model, we can perform the following steps:
1. Remove all generic actions, thus producing Fig. 16.
2. Construct a hierarchal form instead of *contained-in* constructs, thus producing Fig. 17.

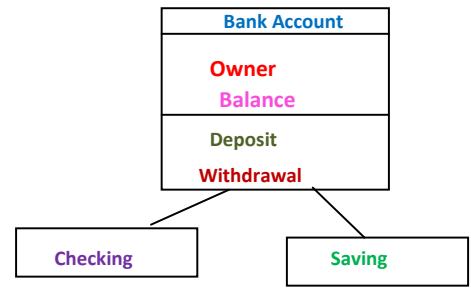

Fig. 17 Hierarchal form of the simplified TM model of the class diagram of the bank account.

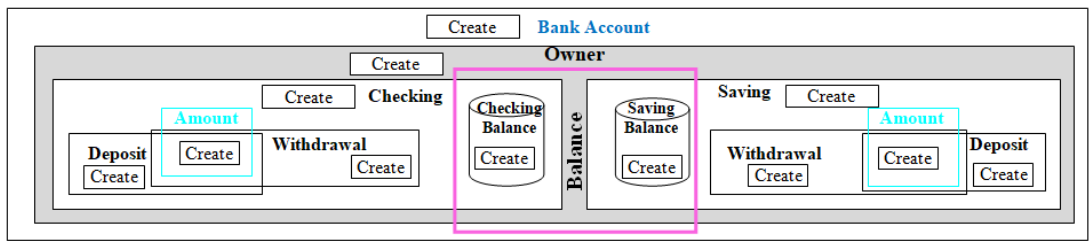

Fig. 16 The static TM model after removing generic actions and flows (arrow).

Note that the simplified form of the static model (Fig. 16) includes create stages that have not been included in Fig. 15 for the sake of simplification. This indicates a two-stage procedure of constructing the bank account system as follows.

1. Using the *create* actions (Fig. 16), an empty template is constructed in a way similar to creating an initial object of a class. For example, the user would create an (empty-null) account, create an (empty) owner, create an (empty) savings account, create an (empty) checking account, etc. This series of creations erects the necessary software and data structure apparatus for an (empty) structure account.

2. Using *process* actions (Fig. 15), the template produced in (1) is filled with values (e.g., the owner, balances, etc.).

However, the TM model is a rich model with further behavioral structures that are applied to Fig. 15 with the following events (see Fig. 18).

Event 1 ($E_1$): Access bank accounts (i.e., not other applications such as loans and credit cards, etc.).
Event 2 ($E_2$): An *Owner* (e.g., identification) is received.
Event 3 ($E_3$): The transaction involves a checking account.
Event 4 ($E_4$): The transaction involves a savings account.
Event 5 ($E_5$): Deposit in a checking account.
Event 6 ($E_6$): Withdrawal from a checking account.
Event 7 ($E_7$): Withdrawal in a savings account.
Event 8 ($E_8$): Deposit in a savings account.
Event 9 ($E_9$): Amount received.
Event 10 ($E_{10}$): In checking account, amount (positive for deposit or negative for withdrawal) flows to be processed.
Event 11 ($E_{11}$): In checking account, balance flows to be processed with amount.
Event 12 ($E_{12}$): In savings account, amount flows to be processed.
Event 13 ($E_{13}$): In savings account, balance flows to be processed with deposit amount.
Event 14 ($E_{14}$): In checking account, a new balance is generated.
Event 15 ($E_{15}$): In savings account, a new balance is generated.
Event 16 ($E_{16}$): In checking account, the balance is updated if a deposit is made.
Event 17 ($E_{17}$): In savings account, the balance is updated if a deposit is made.
Event 18 ($E_{18}$): In checking account, if a withdrawal occurs, a new balance flows to be processed.
Event 19 ($E_{19}$): In savings account, if a withdrawal occurs, a new balance flows to be processed.

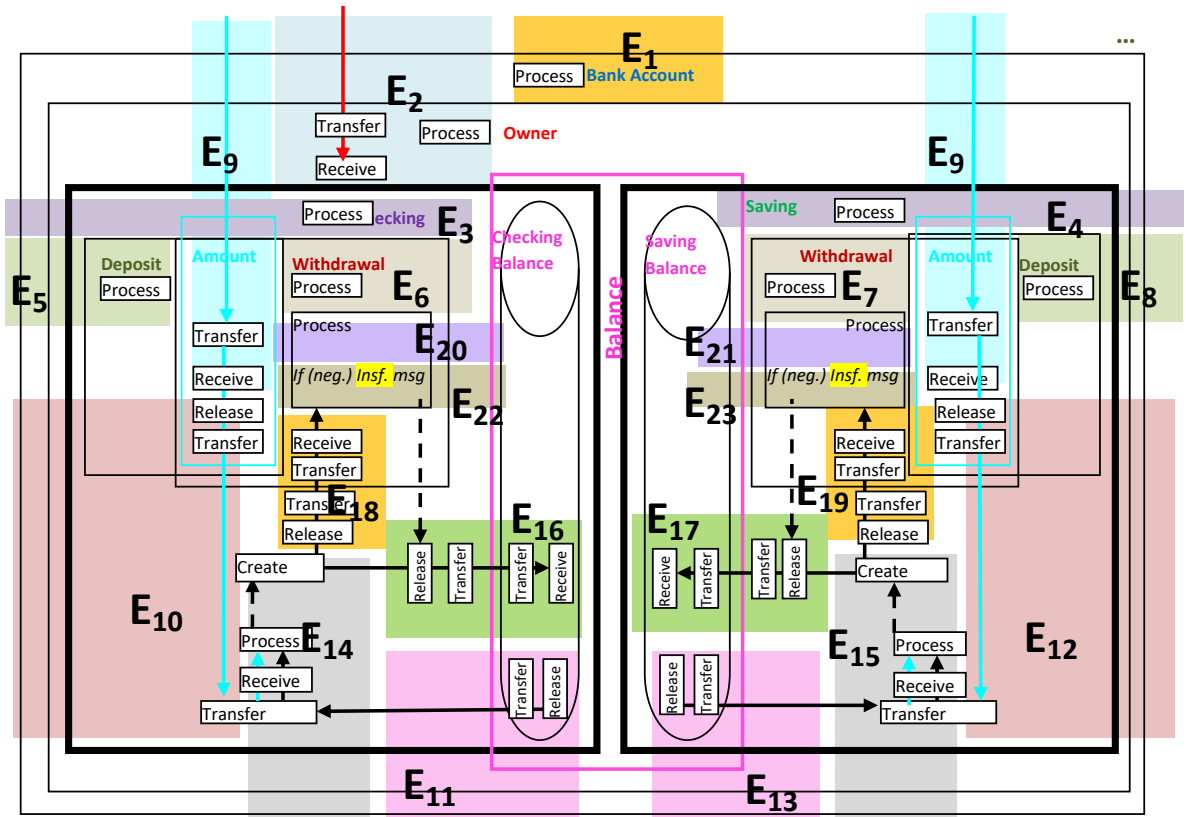

Fig. 18 The TM model of the class diagram of the bank account.

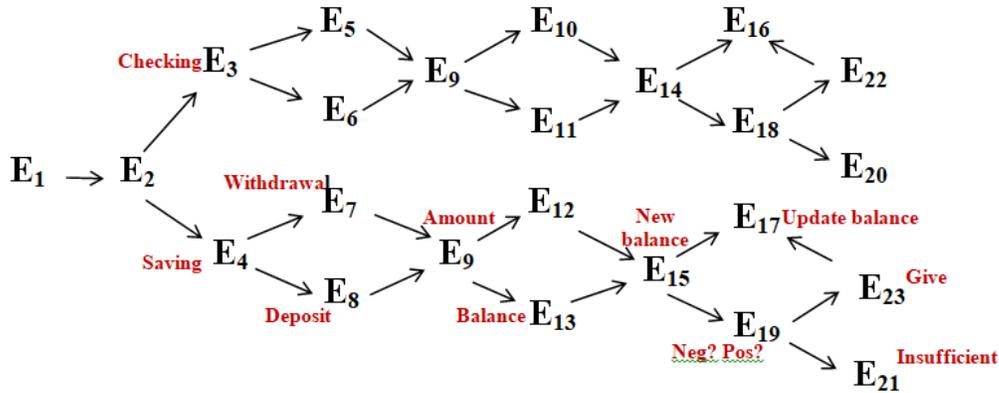

Fig. 19  The behavioral model.

Event 20 ($E_{20}$): In checking account, if there is a withdrawal and the new balance is negative, an "insufficient funds" message is sent.

Event 21 ($E_{21}$): In savings account, if there is a withdrawal and the new balance is negative, an 'insufficient fund' message is sent.

Event 22 ($E_{22}$): In checking account, if there is a withdrawal and the new balance is positive, cash is sent and a balance update is triggered.

Event 23 ($E_{23}$): In savings account, if there is a withdrawal and the new balance is positive, cash is sent and a balance update is triggered.

Fig. 19 shows the behavioral model of this bank account example.

## 5. Further Analysis: Instances of Classes

The analysis in this section aims at further understanding the difference between the two notions of class and object or, in TM terminology, between a thimac and its corresponding time-impregnated version.

The notion of class is generally taken as a formal counterpart of *universals* [36]. However, according to a different conception, properties/attributes are themselves *particulars*, albeit *abstract* ones [41]. Being particular and abstract can be stated as being both properties and objects—things that are typically called *tropes*. Tropes are independent entities whereby universals are resemblance classes of tropes and particular objects are pluralities of tropes. In philosophy, a trope is an instance of a property of a specific entity: the redness of John's T-shirt is a trope that inheres in John's T-shirt [36]. According to Guizzardi et al. [36], "Both John's T-shirt and the redness of John's T-shirt are particulars. However, they are particulars of very different natures. Tropes are particulars which can only exist in other individuals."

In TM modeling, *being red* may mean that the thimac has sharable (universal) flow from redness (transfer, receive) that has particularization in an event (see Fig. 20). In such a view, being red could not be without redness as a source of this being red. On the other hand, it could be claimed no such thing called (universal) *redness* exists. Redness is the result of light reflected in such a way as to be perceived as red. Nevertheless, this trope-like interpretation represents flow in the TM, as shown in Fig. 21. Redness is created independently and locally as a subthimac of the apple thimac. In the case of an apple and grape, two instances of redness resemble each other. Accordingly, the TM model can represent both of these interpretations.

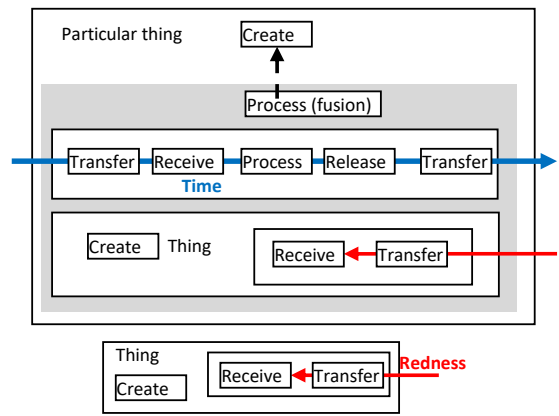

Fig. 20  A thimac of *red (thing)* (bottom) and its particularization/event (top).

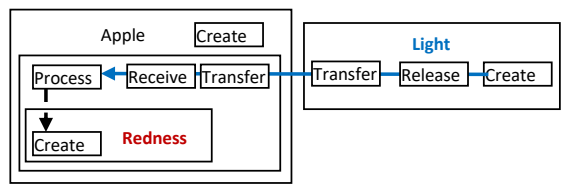

Fig. 21  The TM flow defines the redness of the apple.

Consider the former universality-based TM representation. Thimacs are eventized and thus are converted to particulars. We classify objects (eventized thimacs) as being of the same type. Fig. 22 illustrates the thimac as a general term (top) that is eventized (bottom). As stated previously, the events are represented by their regions. The grape and apple have the same color, which means both have a subthimac of (universal) redness. The "existence" of redness depends on the appearance of the redness thimac in our model; however, the appearance of a redness thimac implies the existence of a (universal) "source" of redness. Similar particularization can be applied to the trope-based representation.

## 6. Conclusion

A class diagram describes a certain conception of static structure in the system and adopts the notion of relationships among those entities. This paper broadens the understanding of such a construct with the aim of developing broad conceptual modeling fundamentals. A new modeling language called TM modeling was employed in this undertaking. The TM concept provides interlacement between structure (components) and actionality whereby actions infiltrate the attributes as much as the classes. However, the class notion in the context of the UML model involves more complex notions than the simple classes discussed in this paper. Nevertheless, from the current study, we can conclude that the involved class description embeds a far richer sematic construct, as reflected in the corresponding TM representation. In general, the current UML class construct seems to restrain conceptual modeling to a rigid form. Future research involving class-based structures that are more complicated would clarify such conclusion.

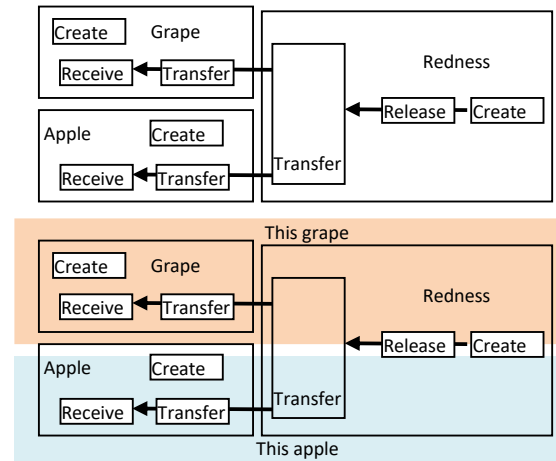

Fig. 22 Grape and apple (classes) being red (top) and *This grape and this apple* (objects) being red (bottom).

## References

[1] Gaines, B.R.: *General systems research: quo vadis?,* In General Systems: Yearbook of the Society for General Systems Research 24, 1–9 (1979)

[2] Guizzardi, G., Wagner, G., Guarino, N., van Sinderen, M.: *An Ontologically Well-Founded Profile for UML Conceptual Models*. In: Persson, A., Stirna, J. (eds.) Advanced Information Systems Engineering. CAiSE 2004. LNCS, vol. 3084, pp. 112–126. Springer, Berlin, Heidelberg (2004)

[3] Dahchour, M., Pirotte, A., Zimányi, E.: *Materialization and Its Metaclass Implementation*. IEEE Trans. on Knowledge and Data Engineering 14(5), 1078–1094 (2002)

[4] Wand, Y., Woo, C., Wand, O.: *Role and Request Based Conceptual Modeling: A Methodology and a CASE Tool.* In: Li, Q., Spaccapietra, S., Yu, E., Olivé, A. (eds.) Conceptual Modeling - ER 2008. LNCS, vol. 5231, pp. 540–541. Springer, Berlin, Heidelberg (2008)

[5] Motik, B., Maedche, A., Volz, R.: *A Conceptual Modeling Approach for Semantics-Driven Enterprise Applications*. In: Meersman, R., Tari, Z. (eds.) On the Move to Meaningful Internet Systems, LNCS, vol. 2519, pp. 1082–1099. Springer, Berlin, Heidelberg (2002)

[6] Fedora: *Object-Oriented Programming*. In: Fedora Project. https://docs.fedoraproject.org/en-US/Fedora/14/html/Musicians_Guide/sect-Musicians_Guide-SC-Basic_Programming-Object_Oriented-Object-Oriented_Programming.html

[7] Pedroni, M., Meyer., B.: *Object-Oriented Modeling of Object-Oriented Concepts: A Case Study in Structuring an Educational Domain*. In: Proceedings of Teaching Fundamental Concepts of Informatics, 4th International Conference on Informatics in Secondary Schools - Evolution and Perspectives, ISSEP 2010, Zurich, Switzerland, January 13-15, 2010. LNCS, vol. 5941, pp. 155–169. Springer, (2010)

[8] OMG, *Unified Modeling Language* [Online]. Available: http://www.uml.org [Accessed: Sept. 24, 2010].

[9] Nikiforova, O., Sejans, J., Cernickins, A.: *Role of UML Class Diagram in Object-Oriented Software Development*. Applied Computer Systems 44, 65–74 (2011)

[10] Fowler, M.: *UML Distilled: A Brief Guide to the Standard Object Modeling Language*, 3rd Edition, Addison-Wesley Professional, (2003)

[11] Bennedsen, J., Caspersen, M.E., Kölling, M.: *Reflections on the Teaching of Programming*. Springer, Berlin/Heidelberg (2008)

[12] Schreiber, G.: *Some Challenge Problems for the Web Ontology Language*. In: University of Amsterdam (no date) http://www.cs.man.ac.uk/~horrocks/OntoWeb/SIG/challenge-problems.pdf

[13] Sedrakyan, G., Poelmans, S., Snoeckc, M.: *Assessing the Influence of Feedback-Inclusive Rapid Prototyping on Understanding the Semantics of Parallel UML Statecharts by Novice Modellers*. Information and Software Technology 82, 159–172 (2017)

[14] Dobing, B., Parsons, J.: *Dimensions of UML Diagram Use: A Survey of Practitioners*. IGI Global, CITY (2008)

[15] Burton-Jones, A., Meso, P.: *Conceptualizing Systems for Understanding: An Empirical Test of Decomposition Principles in Object-oriented Analysis*. Information Systems Research, 17(1), 101–114 (2006)

[16] Fowler, M., Scott, K.: *UML Distilled: A Brief Guide to the Standard Object Modeling Language*, 2nd Edition. Addison-Wesley Prof., Reading, Massachusetts (1999)
[17] Miles, R., Hamilton, K.: *Learning UML 2.0*, 1st Edition. O'Reilly Media, Sebastopol, California (2006)
[18] Halpin, T.: *Fact-Orientation before Object-Orientation (Part 1): The Case for Data Use Cases*. Business Rules Community Newsletter (1999) http://www.brcommunity.com/a1999/a430.html
[19] Hay, D.C.: *Object Orientation and Information Engineering: UML.* In: Reiner, R.S., The Data Administration Newsletter, no. 9 (June 1999), article 5242 at www.tdan.com
[20] Knapp, A., Mossakowski, T.: *Multi-view Consistency in UML: A Survey.* In: Graph Transformation, Specifications, and Nets. LNCS, vol. 10800, pp. 37–60. Springer, Cham (2018)
[21] Al-Fedaghi, S.: *Diagrammatic Formalism for Complex Systems: More than One Way to Eventize a Railcar System*. International Journal of Computer Science and Network Security (IJCSNS) 21(2), 130–141 (2021)
[22] Al-Fedaghi, S.: *UML Modeling to TM Modeling and Back*. International Journal of Computer Science and Network Security (IJCSNS) 21(1), 84–96 (2021)
[23] Al-Fedaghi, S.: *Advancing Behavior Engineering: Toward Integrated Events Modeling*. International Journal of Computer Science and Network Security (IJCSNS) 20(12), 95–107 (2020)
[24] Al-Fedaghi, S.S., BehBehani, M.: *Thinging Machine Applied to Information Leakage*. International Journal of Advanced Computer Science and Applications (IJACSA) 9(9), (2018)
[25] Al-Fedaghi, S., Alrashed, A.: *Threat Risk Modeling*. In: Second International Conference on Communication Software and Networks, Singapore, pp. 405–411, 26-28, Feb. 20 (2010)
[26] Al-Fedaghi, S., Fiedler, G., Thalheim, B.: *Privacy Enhanced Information Systems*. The 15th European-Japanese Conference on Information Modeling and Knowledge Bases: Tallinn, Estonia, pp. 94-111, 2005.
[27] Al-Fedaghi, S.: *Conceptual Temporal Modeling Applied to Databases*, Int. J. Adv. Comput. Sci. Appl. 12(1), 524–534 (2021)
[28] Al-Fedaghi, S.: *UML Modeling to TM Modeling and back*. International Journal of Computer Science and Network Security (IJCSNS) 12(1), 84–96 (2021)
[29] Al-Fedaghi, S., AlSaraf, M.: *High-Level Description of Robot Architecture*. Int. J. Adv. Comput. Sci. Appl. 11(10), 258–267 (2020)
[30] Al-Fedaghi, S.: *Conceptual Software Engineering Applied to Movie Scripts and Stories*. J. Comput. Sci. Technol. 16(12), 1718–1730 (2020)
[31] Al-Fedaghi, S.: *Modeling in Systems Engineering: Conceptual Time Representation*. International Journal of Computer Science and Network Security (IJCSNS) 21(3), 153–164 (2021)
[32] Pirotte, A., Massart, D.: *Integrating Two Descriptions of Taxonomies with Materialization*. Journal of Object Technology 3(5), 143–149 (2004)
[33] Williams, D.C.: *On the Elements of Being II*. Review of Metaphysics 7(2), 171–192 (1953)
[34] Haugen, Ø., Husa, K.E., Runde, R.K., Stølen, K.: *Why Timed Sequence Diagrams Require Three-Event Semantics*. In: Leue, S., Systä, T.J. (eds.) Scenarios: Models, Transformations and Tools. LNCS, vol. 3466, pp 1–25. Springer, Berlin, Heidelberg (2005)
[35] Ehrich, H.D., Goguen, J.A., Sernadas, A.: *A categorical Theory of Objects as Observed Processes*. In: de Bakker, J.W., de Roever, W.P., Rozenberg, G. (eds.) Foundations of Object-Oriented Languages. REX 1990. LNCS, vol. 489, pp. 203–228. Springer, Berlin, Heidelberg (1991)
[36] Guizzardi, G., Masolo, C., Borgo, S.: *In Defense of a Trope-Based Ontology for Conceptual Modeling: An Example with the Foundations of Attributes, Weak Entities and Datatypes*. In: Embley, D.W., Olivé, A., Ram, S. (eds.) Conceptual Modeling - ER 2006. LNCS, vol. 4215, pp. 112–125. Springer, Berlin, Heidelberg (2006)
[37] Pearson (Web site): *A Picture Can Save a Thousand Words: UML Class Diagrams and Java*. In: Inform IT site (Aug 30, 2002) https://www.informit.com/articles/article.aspx?p=29038
[38] Steinhart, E.: *Computational Monadology*. ResearcGate (1999)
[39] Bell, D.: *The Class Diagram: An Introduction to Structure Diagrams in UML 2*. In: IBM Developer (September 15, 2004) https://developer.ibm.com/technologies/web-development/articles/the-class-diagram/
[40] Tutorials, *The Oracle and Java: Object-Oriented Programming Concepts*. (n.d., accessed May, 11, 2021) https://docs.oracle.com/javase/tutorial/java/concepts/
[41] Francesco, O., Paoletti, M.P.: *Properties*. In: Zalta, E.N. (ed.) The Stanford Encyclopedia of Philosophy (Winter 2020 Edition), https://plato.stanford.edu/archives/win2020/entries/properties/